\documentclass[aps,prl,amsmath,amssymb,twocolumn,nofootinbib,showpacs]{revtex4}
\usepackage{pslatex,graphicx,dcolumn,bm,natbib}
\date{\today}

\begin{document}

\title{The Jamming Transition in Granular Systems}

\author{T. S. Majmudar$^1$, M. Sperl$^1$, S. Luding$^2$, R.P. Behringer$^1$\\
\normalsize{$^1$Duke University, Department of Physics, Box 90305, 
Durham, NC 27708, USA,}\\
\normalsize{$^2$Technische Universiteit Delft, DelftChemTech,
Particle Technology, Nanostructured Materials, 
Julianlaan 136, 2628 BL Delft, The Netherlands}
}

\begin{abstract}
Recent simulations have predicted that near jamming for collections of
spherical particles, there will be a discontinuous increase in the
mean contact number, $Z$, at a critical volume fraction, $\phi_c$.
Above $\phi_c$, $Z$ and the pressure, $P$, are predicted to increase as
power laws in $\phi- \phi_c$.  In experiments using photoelastic disks
we corroborate a rapid increase in $Z$ at $\phi_c$ and power-law
behavior above $\phi_c$ for $Z$ and $P$.  Specifically we find
power-law increase as a function of $\phi-\phi_c$ for $Z-Z_c$ with an
exponent $\beta$ around 0.5, and for $P$ with an exponent $\psi$
around 1.1.  These exponents are in good agreement with simulations. We
also find reasonable agreement with a recent mean-field theory for
frictionless particles.

\end{abstract}


\maketitle

A solid, in contrast to a fluid, is characterized by mechanical
stability that implies a finite resistance to shear and isotropic
deformation. While such stability can originate from long-range
crystalline order, there is no general agreement on how mechanical
stability arises for disordered systems, such as molecular and
colloidal glasses, gels, foams, and granular packings \cite{Liu2001}.
For a granular system in particular, a key question concerns how
stability occurs when the packing fraction, $\phi$, increases from
below to above a critical value $\phi_c$ for which there are just
enough contacts per particle, $Z$, to satisfy the conditions of
mechanical stability. In recent simulations on frictionless systems it
was found that $Z$ exhibits a discontinuity at $\phi_c$ followed by a
power law increase for $\phi > \phi_c$
\cite{Silbert2002,OHern2002,OHern2003,Donev2005}.  The pressure is
also predicted to increase as a power-law above $\phi_c$.

A number of recent theoretical studies address jamming, and we
note work that may be relevant to granular systems.  Silbert, O'Hern
et al. have shown in computer simulations of frictionless particles
\cite{Silbert2002,OHern2002,OHern2003} that: a) for increasing $\phi$,
$Z$ increases discontinuously at the transition point from zero to a
finite number, $Z_c$, corresponding to the isostatic value (needed for
mechanical stability); b) for both two- and three-dimensional systems,
$Z$ is expected to continue increasing as $(\phi - \phi_c)^{\beta}$
above $\phi_c$, where $\beta = 0.5$; c) the pressure, $P$, is expected
to grow above $\phi_c$ as $(\phi - \phi_c)^{\psi}$, where $\psi =
\alpha_f -1$ in the simulations, and $\alpha_f$ is the exponent for
the interparticle potential. More recent simulations by Donev et.
al. for hard spheres in three dimensions found a slightly higher value
for $\beta$, $\beta \approx 0.6$, in maximally random jammed packings
\cite{Donev2005}. It is interesting to note that a model for foam 
exhibits quite similar behavior for $Z$ \cite{Durian1995}.
Henkes and Chakraborty \cite{Henkes2005}
constructed a mean field theory of the jamming transition in 2D based
on entropy arguments.  These authors predict power-law scaling for
$P$ and $Z$ in terms of a variable $\alpha$, which is the pressure
derivative of the entropy. By eliminating $\alpha$, one obtains an
algebraic relation between $P$ and $Z - Z_c$ from these predictions,
which we present below in the context of our data. 

While the simulations agree among themselves at least qualitatively, so 
far, these novel features have not been identified in experiments.  
Hence, it is crucial to test these predictions experimentally.  In the 
following, we present experimental data for $Z$ and $P$ vs. $\phi$, based 
on a method that yields accurate determination of the of contacts and 
identifies power laws in $Z$ and $P$ for a two-dimensional experimental 
system of photoelastic disks.  By measuring both $P$ and $Z$, we can also 
obtain a sharper value for the critical packing fraction $\phi_c$, for the 
onset of jamming, and we can test the model of Henkes and Chakraborty.

The relevant simulations have been carried out predominantly for 
frictionless particles.  For real frictional particles there will clearly 
be some differences.  For instance, in the isostatic limit, $Z$ equals 4 
for frictionless disks, whereas for frictional disks, $Z$ is around 3, 
depending on the system details \cite{Alexander1998}. Other predictions 
such as specific critical exponents may also need modification.  However, 
one might hope that the observed experimental behavior, in particular 
critical exponents, might be similar to that for frictionless particles if 
the frictional forces are typically small relative to the normal forces. 
Indeed, in recent experiments, the typical inter-grain frictional forces 
in a physical granular system were found to be {only} about 10\% of the 
normal forces \cite{TSM2005}.

\begin{figure}
\centerline{\includegraphics[width=\columnwidth]{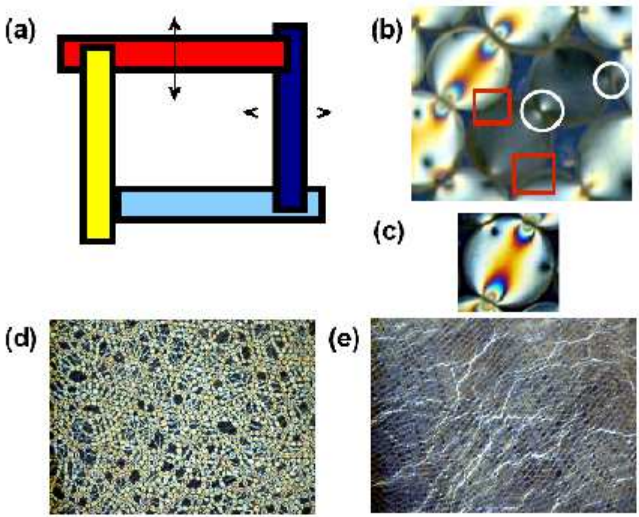}}
\caption{\label{f_exp} (a) Schematic cross-section of biaxial cell
experiment (not to scale).  Two walls can be moved independently to
obtain a desired sample deformation.  (b) Examples of contacts and
particles that are either close but not actually in contact, or
contacts with very small forces. Circles show true contacts, squares
show false apparent contacts.  (c) Image of a single disk at the
typical resolution of the experiment. (d) Sample image of highly
jammed/compressed state and (e) almost unjammed state.}
\end{figure}

Fig.~\ref{f_exp}a shows a schematic of the apparatus. We use a
bidisperse mixture (80\% small and 20\% large particles) of
approximately 3000 polymer (PSM-4) photoelastic (birefringent under
stress) disks with diameter 0.74 cm or 0.86 cm. This ratio preserves a
disordered system.  The disks have Young's modulus of 4 MPa, and a
static coefficient of friction of 0.85.  The model granular system is
confined in a biaxial test cell (42cm$\times$42cm with two movable
walls) which rests on a smooth Plexiglas sheet. The displacements of
the walls can be set very precisely with stepper motors.  The linear
displacement step size used in this experiment is $40 \mu$m, which is
approximately $0.005D$, where $D$ is the average diameter of the disks.
The deformation $\delta$ per particle is less than 1\% in the
compressed state.  The setup is horizontal and placed between crossed
circular polarizers.  It is imaged from above with an 8 MP CCD color
camera which captures roughly 1200 disks in the center of the cell,
enabling us to visualize the stress field within each disk
(Fig.~\ref{f_exp}).  We then obtain good measurements of the vector
contact forces (normal and tangential = frictional components)
\cite{TSM2005}.

We also use the particle photoelasticity to accurately determine the
presence or absence of contacts between particles. In numerical
studies one can use a simple overlap criterion to determine contacts:
a contact occurs if the distance between particle centers is smaller
than the sum of the particle radii. However, in experimental systems,
a criterion based solely on the particle centers is susceptible to
relatively large errors which include false positives
(Fig.~\ref{f_exp}b - squares) as well as false negatives (circles). As
seen in Fig.~\ref{f_exp}b, the contacts through which there is force
transmission appear as source points for the stress pattern. Further 
details are given in the supplementary material, section~I.

We use two protocols to produce different packing fractions: we either
compress the system from an initially stress-free state, or decompress
the system until the end state is essentially a stress-free state. The
results for both protocols are the same within error bars above
$\phi_c$; below jamming, the data for $Z$ obtained by compression are
a few percent below those for decompression. Below, we will present
decompression data. Figures~\ref{f_exp}d,e show the initial highly
stressed state and the end state after decompression,
respectively. After each decompression step, we apply tapping to relax
stress in the system.  This could be seen as roughly analogous to the
annealing process invoked in some simulations.  Two images are captured at
each state: one without polarizers to determine the disk centers and
one with polarizers to record the stress.

The average $Z$ can be computed either by counting only the force
bearing disks or by counting all the disks including rattlers which do
not contribute to the mechanical stability of the system. We consider
as rattlers, all the disks which have less than 2 contacts. For the
number of rattlers beyond the transition point we find an exponential
decrease with $\phi-\phi_c$; hence, a divergence in the number of
rattlers at $\phi_c$ is not indicated by the data.

We next compute the Cauchy stress tensor for each disk, $\sigma_{ij} = 
\frac{1}{2A} \sum{(F_{i}x_{j} + F_{j}x_{i} )} $; $P$, is the trace of this 
tensor.  Here, $A$ is the Voronoi area for the given disk, and the sum is 
taken over contacts for a given disk.  We then compute the average of the 
pressure over the ensemble of disks in the system.  For the data presented 
below, we performed two sets of experiments: one with a larger range, 
$0.8390 \leq \phi \leq 0.8650$, and also larger step size, $\Delta\phi = 
0.016$, and -- after the jamming region was identified -- a second set at 
a finer scale with $0.840745 \leq \phi \leq 0.853312$, with a step size, 
$\Delta\phi = 0.000324$.

\begin{figure}
\centerline{\includegraphics[width=\columnwidth]{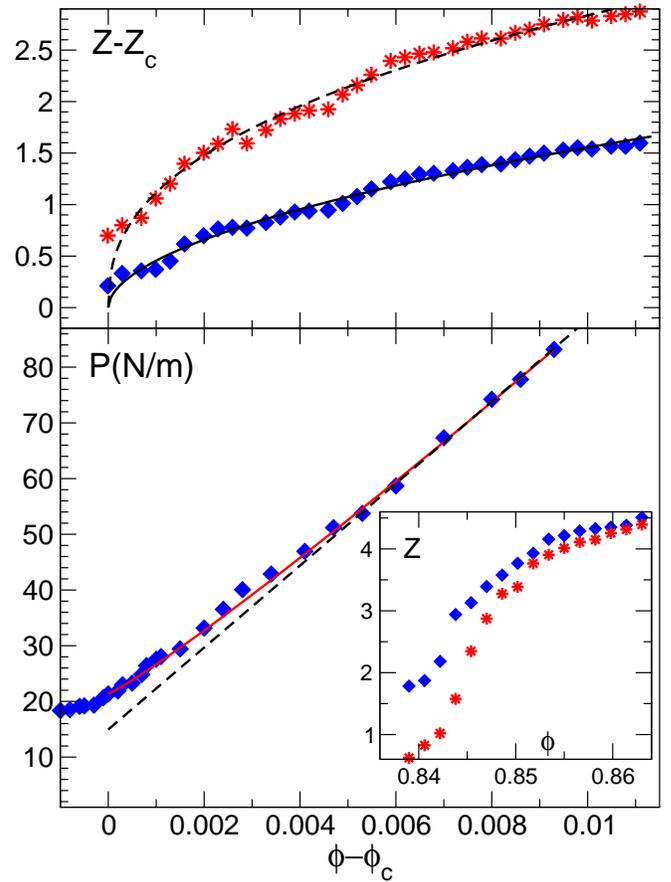}}
\caption{\label{f_ZPZ}Average contact number and pressure at the
jamming transition. Top and bottom panels show $Z-Z_c$ and $P$ vs
$\phi-\phi_c$, respectively, with rattlers included (stars) or
excluded (diamonds). Dashed and full curves in the top panel give
power-law fits $(\phi- \phi_c)^\beta$ with $\beta = 0.495$ and 0.561
for the case with and without rattlers, respectively. Full curve in
the lower panel gives the fit $(\phi-\phi_c)^\psi$ with $\psi=1.1$;
dashed line shows a linear law for comparison. Inset: $Z$ vs $\phi$
for a larger range in $\phi$.}
\end{figure}

The inset in Fig.~\ref{f_ZPZ} shows data for $Z$ over a broad range of 
$\phi$ (with rattlers--stars; without--squares).  These data show a 
significant rise in $Z$ at the jamming transition. While this rise is not 
sharply discontinuous, it occurs over a very small range in $\phi$.  At 
higher $\phi$, the variations of the curves are similar with and without 
rattlers. At lower $\phi$, their behavior differs: The values of Z drop 
lower for the case with rattlers. The pressure $P(\phi - \phi_c)$ in 
Fig.~\ref{f_ZPZ} shows a flat background below jamming, and then a sharp 
positive change in slope at a well defined $\phi$. The pressure is not 
identically zero below jamming for similar reasons that the jump in $Z$ is 
not perfectly sharp, as discussed below.

To compare these experimental results to predictions above $\phi_c$, we 
carry out least squares fits of $Z-Z_c$ and $P$ to $\phi-\phi_c$. These 
fits depend on the choice of $\phi_c$, which has some ambiguity due to the 
rounding; the data allow a range from around 0.840 to 0.843.  In fact, 
$\phi_c$ can be determined in several ways: the point where $Z$ reaches 3, 
the point where $P$ begins to rise above the background, etc. (cf. 
supplemetary material). We show results of these fits in Fig.~\ref{f_ZPZ}, 
starting with the upper panel, which shows power-law fits $(Z-Z_c) \propto 
(\phi-\phi_c)^{\beta}$.  The fitted exponent $\beta$ depends on the choice 
of $\phi_c$ but the variation is small without rattlers, $0.494 \leq \beta 
\leq 0.564$, and somewhat larger with rattlers, $0.363 \leq \beta \leq 
0.525$. The details for several different specific fits are given in the 
supplementary material, section~II. The point $\phi_c = 0.84220$ where $P$ 
rises above the background level is used in Fig.~\ref{f_ZPZ}, and yields a 
consistent fit for both $P$ and $Z$. The point where $Z$ reaches 3 for the 
case without rattlers agrees with the previous case to within $\delta 
\phi_c = 0.0005$, and the exponents are quite similar. Comparing with the 
simulations for frictionless particles, we find that our values of $\beta 
\approx 0.55$ for the data without rattlers are larger than the value of 
0.5 reported in \cite{Silbert2002,OHern2002}, but smaller than those of 
Donev et.\ al.\ \cite{Donev2005} obtaining 0.6 in 3D. In contrast, for a 
model of frictional disks under shear, Aharonov and Sparks 
\cite{aharonov1999} obtain the much lower value of 0.36. However, a direct 
comparison is not possible to the present case of jamming under isotropic 
conditions.

Figure~\ref{f_ZPZ} shows the variation of $P$ with $\phi$ in the lower
panel, indicating a clear transition at $\phi_c = 0.8422 \pm
0.0005$. For this choice of $\phi_c$, $P$ increases as $P \propto
(\phi - \phi_c)^\psi$ with $\psi = 1.1 \pm 0.05$ above $\phi_c$.  This
value of $\psi$ pertains to a fit over the full range $\phi\geq
\phi_c$ of Fig.~\ref{f_ZPZ}; a larger exponent would be obtained if
the fit range were limited to very close to $\phi_c$.  This value is
close to the value $\psi = 1.0$ found \cite{Silbert2002,OHern2002} for
a linear force law, and this linear law is indicated as a dashed line
in Fig.~\ref{f_ZPZ}. One expects such a linear force law (with a
logarithmic correction) for ideal disks, but direct mechanical
calibration of the force law for the cylinders is closer to
$\delta^{3/2}$ (see supplementary material).  This rather high
exponent for the force law is attributable to the small asperities,
which influence the force law for small deformations.  However, the
photoelastic response is detectable only for $\delta > 150\mu$m, and
for such $\delta$'s, the force law is close to locally linear in
$\delta$.

From the $P$ vs. $\phi$ data, we can also obtain the bulk modulus, $B = 
-A\partial P/\partial A$, where $A$ is the area enclosed by the system 
boundaries.  Since, $\phi = A_p/A$, where $A_p$ is the (presumably fixed) 
area occupied by the disks, $B = \phi \partial P/\partial \phi$.  Then, $B 
\propto (\phi -\phi_c)^{\psi -1}$, which gives a weak pressure variation 
of $B$ above $\phi_c$.  We note that {anomalous} results for the bulk 
modulus have been observed in {acoustical} experiments by Jia, and 
discussed by Makse et.\ al.\ \cite{makse99}, where the bulk modulus near 
$\phi_c$ varied faster with $P$ than was previously expected because of 
changes in $Z$.

\begin{figure}
\centerline{\includegraphics[width=\columnwidth]{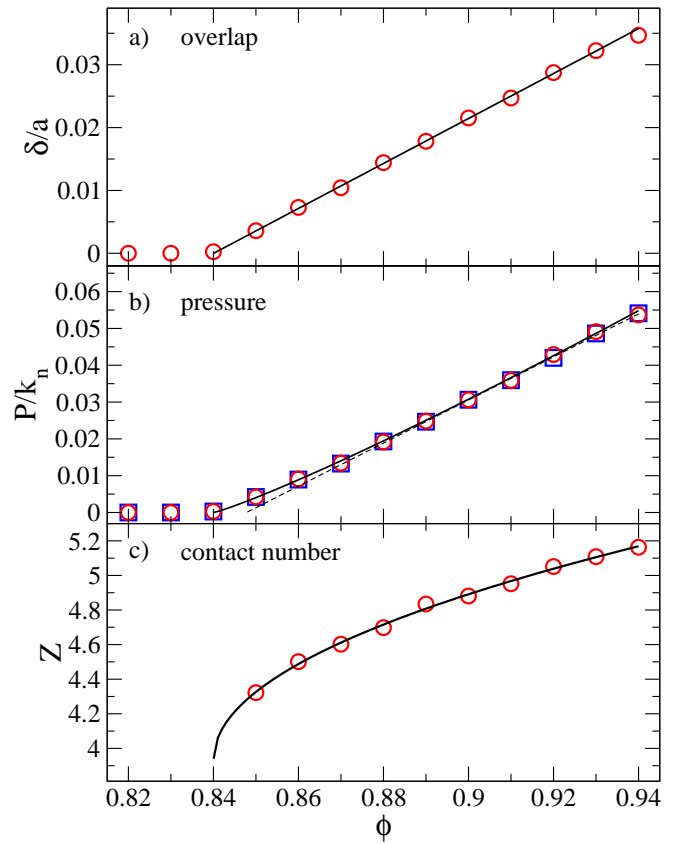}}
\caption{\label{MD} Results from new computer simulations. For all
fits $\phi_c=0.84005$. (a) Average overlap per particle in units of
the mean particle radius is linear in $\phi-\phi_c$. (b) $P$ obtained
from the Cauchy stress tensor (circles) and the force on the walls
(squares) satisfy a power law $(\phi-\phi_c)^\Psi$ with $\psi=1.13$;
dashed line shows a linear law for comparison. (c) $Z$ (rattlers
included) exhibits a power law $Z-Z_c\propto(\phi- \phi_c)^\beta$ with
$Z_c=3.94$ and $\beta=0.5015$.}
\end{figure}

Since $P$ in Fig.~\ref{f_ZPZ} corresponds closely to expectations
for a linear force law, we performed a computer simulation for a
polydisperse system of 1950 particles with a linear force law
($k_n=10^5$N/m) without friction; details can be found in
\cite{Madadi2004}. In Fig.~\ref{MD} the results are shown for a larger
range in density than done in earlier studies. All the data in
Fig.~\ref{MD} can be fitted with a single value for the transition
density of $\phi_c = 0.84005$. While the average overlap per particle
(equivalent of the deformation $\delta$ for physical particles) is
clearly linear in $\phi$, the pressure $P$ is not: $P$ increases faster than
linear with an exponent close to the one found in the experiment. $Z$
is also consistent with a power-law exponent close to 0.5. With the
rattlers included, $Z$ at $\phi_c$, $Z_c=3.94$, is slightly below the
isostatic value of 4 for a frictionless system of disks.

\begin{figure}
\centerline{\includegraphics[width=\columnwidth]{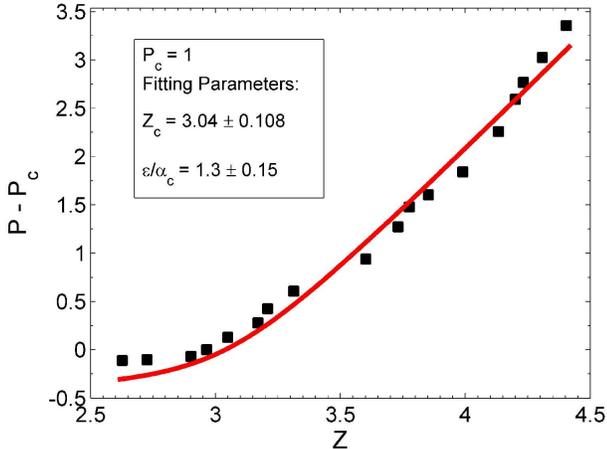}}
\caption{\label{Bulbulfit} Pressure vs. Z; Experimental data and a fit to 
the model of Henkes and Chakraborty \cite{Henkes2005}.  In this fit, the 
constant $C$ defined in the text is treated as an adjustable parameter. 
The other fitting parameter is $Z_c$.}
\end{figure}

To connect with the predictions of Henkes and Chakraborty
\cite{Henkes2005}, we consider $P - P_c$ vs. $Z$.  The prediction from
their Eq.~(10) is equivalent to $(P -P_c)/P_c = u - [(4u^2 + 1)^{1/2}
-1]/2$, where $u = C(Z - Z_c)$ and $C = \epsilon/\alpha_c$ is a
system-dependent constant.  Thus, $\epsilon$ is a measure of the grain
elasticity, and $\epsilon = 0$ corresponds to completely rigid grains.
Also, $\alpha_c$ is the critical value for $\alpha$.  In fitting to
this form, we may adjust $P_c$, (within reason) $Z_c$, and $C$.  In
Fig.~\ref{Bulbulfit} we find reasonable although not perfect agreement
with this prediction (above $\phi_c$), and obtain $Z_c = 3.04$, which
is close to the isostatic value $Z_c = 3$.

We now turn to the rounding that we observe in $Z$ quite close to the 
transition, and the background pressure that we obtain near $\phi_c$. One 
possible explanation is the friction between the disks and the Plexiglas 
base.  This could help freeze in contact forces and contacts.  However, a 
simple estimate of the upper bound for the friction with the base shows 
that this cannot be a significant effect, at least as regards the pressure 
background. To obtain an estimated upper bound for the base friction on 
$P$, we assume that base friction can support inter-grain contact forces 
corresponding to the maximum base frictional force per grain, $F_f = 
\mu_{ba} m g = 2.8 \times 10^{-3}$~N, where $m$ is the mass of a grain and 
$\mu_{ba} < 1$ is the friction between a particle and the base. Assuming 
$Z$ inter-particle contacts and one particle-base contact per grain, we 
estimate the resulting upper bound on the perturbation to the pressure as 
$\delta P \simeq (Z F_f R)/(\pi R^2) \simeq 0.22Z$~N/m, where $R$ is a 
disk radius.  Since $Z \simeq 3$, this pressure is almost two orders of 
magnitude too small to be of relevance.  An additional issue concerns the 
anisotropy that is induced during compression or expansion by the 
apparatus.  This induced anisotropy is difficult to avoid and/or relax 
close to $\phi_c$ even in the simulation, but it remains small. It is 
visible in Fig.~\ref{f_exp}e, where a weak array of force chains tends to 
slant from lower left to upper right. Among other reasons, the anisotropy 
can be induced by wall friction due to the confining lateral boundaries of 
the biaxial apparatus.

We conclude by noting that these experiments, the first of which we
are aware, demonstrate the critical nature of jamming in a real
granular material.  Our results take advantage of the high accuracy in
contact number $Z$ that is afforded when the particles are photoelastic.
$Z$ shows a very rapid rise at a packing density $\phi_c = 0.8422$. The 
fine resolution in density allows us to see that the transition is not 
as sharply discontinuous under the present experimental conditions
as in the computer simulation. Above $\phi_c$, $Z$ and $P$ follow power laws 
in $\phi - \phi_c$ with respective exponents $\beta$ of 0.5 to 0.6 and
$\psi \approx 1.1$.  The values for both $\beta$ and $\psi$ are
consistent with recent simulation results for {\em frictionless} particles.
In addition, we find reasonable agreement with a mean field model of
the granular jamming transition, again for frictionless particles.
These results suggest that effects of friction on jamming are likely
modest, although perhaps not ignorable.  That jamming in the experiment 
occurs over a narrow, but finite range in $\phi$ seems mostly to be
caused by small residual shear stresses that are induced by
interactions with the walls confining the sample (not the base
supporting the particles).  The ability of a small amount of shear to
affect the jamming transition is interesting, and points to the need
for a deeper understanding of the effects of anisotropy.

This work was supported by NSF-DMR0137119, NSF-DMR0555431, 
NSF-DMS0244492, the US-Israel Binational Science Foundation \#2004391, and 
DFG-SP714/3-1.
We thank E. Aharonov, B. Chakraborty, D. J. Durian, M. van Hecke, 
C.S. O'Hern, and S. Torquato for helpful discussions.

\bibliographystyle{apsrev}

\begin{thebibliography}{10}
\expandafter\ifx\csname natexlab\endcsname\relax\def\natexlab#1{#1}\fi
\expandafter\ifx\csname bibnamefont\endcsname\relax
  \def\bibnamefont#1{#1}\fi
\expandafter\ifx\csname bibfnamefont\endcsname\relax
  \def\bibfnamefont#1{#1}\fi
\expandafter\ifx\csname citenamefont\endcsname\relax
  \def\citenamefont#1{#1}\fi
\expandafter\ifx\csname url\endcsname\relax
  \def\url#1{\texttt{#1}}\fi
\expandafter\ifx\csname urlprefix\endcsname\relax\def\urlprefix{URL }\fi
\providecommand{\bibinfo}[2]{#2}
\providecommand{\eprint}[2][]{\url{#2}}

\bibitem[{\citenamefont{Liu and Nagel}(2001)}]{Liu2001}
\bibinfo{author}{\bibfnamefont{A.}~\bibnamefont{Liu}} \bibnamefont{and}
  \bibinfo{author}{\bibfnamefont{S.}~\bibnamefont{Nagel}},
  \emph{\bibinfo{title}{Jamming and Rheology: Constrained Dynamics on
  Microscopic and Macroscopic Scales}} (\bibinfo{publisher}{Taylor {\&}
  Francis}, \bibinfo{address}{New York}, \bibinfo{year}{2001}).

\bibitem[{\citenamefont{Silbert et~al.}(2002)\citenamefont{Silbert, Erta\c{s},
  Grest, Halsey, and Levine}}]{Silbert2002}
\bibinfo{author}{\bibfnamefont{L.~E.} \bibnamefont{Silbert}},
  \bibinfo{author}{\bibfnamefont{D.}~\bibnamefont{Erta\c{s}}},
  \bibinfo{author}{\bibfnamefont{G.~S.} \bibnamefont{Grest}},
  \bibinfo{author}{\bibfnamefont{T.~C.} \bibnamefont{Halsey}},
  \bibnamefont{and} \bibinfo{author}{\bibfnamefont{D.}~\bibnamefont{Levine}},
  \bibinfo{journal}{Phys.~Rev.~E} \textbf{\bibinfo{volume}{65}},
  \bibinfo{pages}{031304} (\bibinfo{year}{2002}).

\bibitem[{\citenamefont{O'Hern et~al.}(2002)\citenamefont{O'Hern, Langer, Liu,
  and Nagel}}]{OHern2002}
\bibinfo{author}{\bibfnamefont{C.}~\bibnamefont{O'Hern}},
  \bibinfo{author}{\bibfnamefont{S.~A.} \bibnamefont{Langer}},
  \bibinfo{author}{\bibfnamefont{A.~J.} \bibnamefont{Liu}}, \bibnamefont{and}
  \bibinfo{author}{\bibfnamefont{S.~R.} \bibnamefont{Nagel}},
  \bibinfo{journal}{Phys.~Rev.~Lett.} \textbf{\bibinfo{volume}{88}},
  \bibinfo{pages}{075507} (\bibinfo{year}{2002}).

\bibitem[{\citenamefont{O'Hern et~al.}(2003)\citenamefont{O'Hern, Silbert, Liu,
  and Nagel}}]{OHern2003}
\bibinfo{author}{\bibfnamefont{C.}~\bibnamefont{O'Hern}},
  \bibinfo{author}{\bibfnamefont{L.~E.} \bibnamefont{Silbert}},
  \bibinfo{author}{\bibfnamefont{A.~J.} \bibnamefont{Liu}}, \bibnamefont{and}
  \bibinfo{author}{\bibfnamefont{S.~R.} \bibnamefont{Nagel}},
  \bibinfo{journal}{Phys.~Rev.~E} \textbf{\bibinfo{volume}{68}},
  \bibinfo{pages}{011306} (\bibinfo{year}{2003}).

\bibitem[{\citenamefont{Donev et~al.}(2005)\citenamefont{Donev, Torquato, and
  Stillinger}}]{Donev2005}
\bibinfo{author}{\bibfnamefont{A.}~\bibnamefont{Donev}},
  \bibinfo{author}{\bibfnamefont{S.}~\bibnamefont{Torquato}}, \bibnamefont{and}
  \bibinfo{author}{\bibfnamefont{F.~H.} \bibnamefont{Stillinger}},
  \bibinfo{journal}{Phys.~Rev.~E} \textbf{\bibinfo{volume}{71}},
  \bibinfo{pages}{011105} (\bibinfo{year}{2005}).

\bibitem{Durian1995}
\bibinfo{author}{D. J. Durian}
\newblock \emph{\bibinfo{journal}{Phys.~Rev.~Lett.}}
  \textbf{\bibinfo{volume}{75}}, \bibinfo{pages}{4780} 
(\bibinfo{year}{1995}).

\bibitem[{\citenamefont{Henkes and Chakraborty}(2005)}]{Henkes2005}
\bibinfo{author}{\bibfnamefont{S.}~\bibnamefont{Henkes}} \bibnamefont{and}
  \bibinfo{author}{\bibfnamefont{B.}~\bibnamefont{Chakraborty}},
  \bibinfo{journal}{Phys.~Rev.~Lett.} \textbf{\bibinfo{volume}{95}},
  \bibinfo{pages}{198002} (\bibinfo{year}{2005}).

\bibitem[{\citenamefont{Alexander}(1998)}]{Alexander1998}
\bibinfo{author}{\bibfnamefont{S.}~\bibnamefont{Alexander}},
  \bibinfo{journal}{Phys. Rep.} \textbf{\bibinfo{volume}{296}},
  \bibinfo{pages}{65} (\bibinfo{year}{1998}).

\bibitem[{\citenamefont{Majmudar and Behringer}(2005)}]{TSM2005}
\bibinfo{author}{\bibfnamefont{T.~S.} \bibnamefont{Majmudar}} \bibnamefont{and}
  \bibinfo{author}{\bibfnamefont{R.~P.} \bibnamefont{Behringer}},
  \bibinfo{journal}{Nature} \textbf{\bibinfo{volume}{435}},
  \bibinfo{pages}{1079} (\bibinfo{year}{2005}).

\bibitem{aharonov1999}
\bibinfo{author}{E. Aharonov} and \bibinfo{author}{D. Sparks}
\newblock {\bibinfo{journal}{Phys.~Rev.~E}} \textbf{\bibinfo{volume}{60}},
  \bibinfo{pages}{6890} (\bibinfo{year}{1999}).

\bibitem[{\citenamefont{Makse et~al.}(1999)\citenamefont{Makse, Gland, Johnson,
  and Schwartz}}]{makse99}
\bibinfo{author}{\bibfnamefont{H.~A.} \bibnamefont{Makse}},
  \bibinfo{author}{\bibfnamefont{N.}~\bibnamefont{Gland}},
  \bibinfo{author}{\bibfnamefont{D.~L.} \bibnamefont{Johnson}},
  \bibnamefont{and} \bibinfo{author}{\bibfnamefont{L.~M.}
  \bibnamefont{Schwartz}}, \bibinfo{journal}{Phys. Rev. Lett.}
  \textbf{\bibinfo{volume}{83}}, \bibinfo{pages}{5070} (\bibinfo{year}{1999}).

\bibitem[{\citenamefont{Madadi et~al.}(2004)\citenamefont{Madadi, Tsoungui,
  L{\"a}tzel, and Luding}}]{Madadi2004}
\bibinfo{author}{\bibfnamefont{M.}~\bibnamefont{Madadi}},
  \bibinfo{author}{\bibfnamefont{O.}~\bibnamefont{Tsoungui}},
  \bibinfo{author}{\bibfnamefont{M.}~\bibnamefont{L{\"a}tzel}},
  \bibnamefont{and} \bibinfo{author}{\bibfnamefont{S.}~\bibnamefont{Luding}},
  \bibinfo{journal}{Int. J. of Solids and Structures}
  \textbf{\bibinfo{volume}{41}}, \bibinfo{pages}{2563} (\bibinfo{year}{2004}).


\end{thebibliography}

\end{document}


\title{The Jamming Transition in Granular Systems -- Supplementary 
Information}

\author{T. S. Majmudar$^1$, M. Sperl$^1$, S. Luding$^2$, R.P. Behringer$^1$\\
\normalsize{$^1$Duke University, Department of Physics, Box 90305, 
Durham, NC 27708, USA,}\\
\normalsize{$^2$Technische Universiteit Delft, DelftChemTech,
Particle Technology, Nanostructured Materials,}\\ 
\normalsize{Julianlaan 136, 2628 BL Delft, The Netherlands}
}

\maketitle

\section{Supplementary methods}

In this supplement, we provide experimental details which we discuss
in the context of Fig.~1.  A key point concerning the experiments is
the use of photoelasticity (stress-induced birefringence) to obtain
vector forces at interparticle contacts.  This technique has the added
advantage of determining to good accuracy whether a contact is present
or not.

\paragraph{Photoelastic Method and Determination of Contacts}

A stressed photoelastic particle (in our case, a disk) when viewed
through crossed circular polarizers, shows a pattern of light and dark
bands. The light rays traversing the polarizers and a particle (along
the axial direction of the disk) have an intensity $I = I_o
sin^2[(\sigma_1-\sigma_2)C]$. Here, the $\sigma_i$ are the principle
stresses within the particle; $C$ is a constant that depends on the
the thickness and properties of the disk, and on the wavelength of the
light \cite{frocht_41}. Given a set of contacts for a disk, and forces
at these contacts, the specific photoelastic pattern is
determined. Here, we take advantage of the fact that a two-dimensional
description for the stresses is appropriate.  Assuming that the
contact forces are well-approximated as point-like, the Boussinesq
solution gives the stresses within the disk \cite{Frocht1948}. For
these experiments, we solve the inverse problem: we have the light
intensities of the photoelastic pattern within a disk, and we find the
contact forces. We use an automated computer algorithm which uses the
vector contact forces as nonlinear least-squares fit parameters.  The
fitting procedure minimizes differences between the experimentally
measured intensity pattern for a disk and the intensity pattern that
would be obtained for a given set of contact forces \cite{TSM2005}.

In order to improve the discrimination between false and true contacts
we employ a two step process. The first step involves obtaining
possible contacts based on the distances between disk centers; if the
particle centers are within $D \pm 0.1 D$, where $D$ is the mean
center-to-center distance of a particle pair, the disks are considered
to be in potential contact.  This estimate of contacts is markedly
improved by utilizing the photoelastic stress images at various
exposure times for each state, such that eventually most of the force
transmitting contacts can be seen.  As seen in Fig.~1b, the contacts
through which there is force transmission, appear as source points for
the stress pattern. This effect can be quantified by measuring the
intensity and the gradient square of the intensity ($G^{2} = |\nabla
I|^2$ where the gradient is taken in the plane of the disk) around the
contact \cite{howell_99}. A true, force bearing contact can be
distinguished by employing appropriate thresholds in intensity and in
$G^{2}$. The thresholds in intensity and in $G^2$ are useful in
capturing contacts with very small forces, since these quantities are
higher near force bearing contacts.  The final error in average $Z$ is
around 3.5\% for rather low $\phi$, and around 1.5\% for higher
$\phi$.

\paragraph{Calibration of the Force Law}

A direct mechanical calibration for the particles using a digital
force gage is shown in Fig.~\ref{calibration}: The dotted curve shows
a force law $F\propto\delta^{3/2}$. A linear fit describes the
calibration data well for $\delta > 250\mu$m which is comparable to
the surface roughness of the cylinders. The photoelastic response is
detectable for displacements that exceed the right end of the gray bar
at $\delta \approx 150\mu$m.  In the effective range for the
photoelastic technique, $\delta > 150\mu$m, the force vs. displacement
curve is reasonably well described by a straight line.

\renewcommand{\thefigure}{S\arabic{figure}}
\begin{figure}[h]
\centerline{\includegraphics[width=\columnwidth]{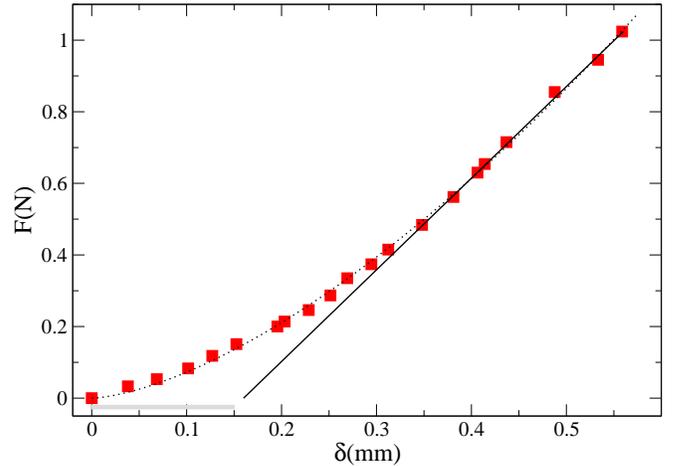}}
\caption{\label{calibration} Calibration of the contact force $F$ for
a representative disk pushed against a hard surface by a displacement
$\delta$. The experimental data (squares) are fitted by the power law
$F = 2.52\text{N}\;(\delta^{1.54})$ (dotted) and by the linear law $F
= 2.56\text{N}\; (\delta - 0.16)$ (full curve). Here, all lengths are
given in mm.  The gray bar indicates the roughness of the cylinder
surface. Photoelastic response is reliably detectable to the right of
this bar.}
\end{figure}

\section{Supplementary Table}
\paragraph{Details of the Fitting Procedure}

\renewcommand{\thetable}{S\Roman{table}}
\begin{table*}
\begin{center}
\begin{tabular}{| c | c | c | c | c | c | c |} 
\cline{1-7}
 \multicolumn{1}{| c |}{$\phi_c$}&\multicolumn{3}{| c  |}{Without Rattlers}&\multicolumn{3}{| c |}{With Rattlers}\\
\cline{1-7}
  &$Z_{c}$&$\beta$&RMSE&$Z_{c}$&$\beta$&RMSE\\
\cline{1-7}
0.84058 &$2.397\pm 0.135$ &$0.517 \pm 0.064$&0.049&$1.198 \pm 0.310$&$0.502 \pm 0.093$&0.109\\
\cline{1-7}
0.84075&$2.512 \pm 0.138$&$0.547 \pm 0.073$&0.051&$1.071 \pm 0.359$&$0.460 \pm 0.090$&0.103\\
\cline{1-7}
0.84172&$2.632 \pm 0.151$&$0.494 \pm 0.077$&0.045&$0.9747 \pm 0.458$&$0.363 \pm 0.083$&0.080\\
\cline{1-7}
0.84204&$2.858 \pm 0.127$&$0.564 \pm 0.086$&0.045&$1.183 \pm 0.413$ &$0.367 \pm 0.079$&0.072\\
\cline{1-7}
0.84220&$2.838 \pm 0.171$&$0.533 \pm 0.102$&0.046&$1.490 \pm 0.427$&$0.405 \pm 0.096$&0.072\\
\cline{1-7}
0.84236&$2.916 \pm 0.133$&$0.556 \pm 0.093$&0.046&$1.744 \pm 0.298$&$0.445 \pm 0.088$&0.075\\
\cline{1-7}
0.84269&$3.003 \pm 0.124$&$0.563 \pm 0.095$&0.043&$1.989 \pm 0.267$&$0.469 \pm 0.092$&0.071\\
\cline{1-7}
0.84301&$3.075 \pm 0.120$&$0.560 \pm 0.095$&0.041&$2.280 \pm 0.235$&$0.525 \pm 0.108$&0.072\\
\cline{1-7}
\end{tabular}
\end{center}
\caption{\label{Table1} Power-law exponents and critical contact numbers 
obtained as fitting parameters, at various critical packing fractions. The 
RMSE gives the root mean squared errors for the fits. The indicated 
uncertainties in both $Z_c$, and $\beta$ are obtained from the 
95\% confidence interval of the best-fit parameter values.}

\end{table*}

For the fit of the data with the power law $Z-Z_c = a(\phi-\phi_c)^\beta$
we examine a range of values for $\phi_c$ and obtain the 
exponents for the power-law fits given in Table \ref{Table1}. Here, 
$\phi_c$ is selected, and $Z_c$, and $\beta$ are the fitting parameters. 
For the case without rattlers, $\beta$ ranges from 0.49 to 0.56, and $Z_c$ 
ranges from 2.40 to 3.08. For the case with rattlers, $\beta$ shows more 
variation (0.36 - 0.52), and the errors in $Z_c$ are larger. For the 
entire range of $\phi$, the root mean squared errors (RMSE) are larger for 
the case with rattlers (0.071 - 0.109), than for the case without rattlers 
(0.041 - 0.051), indicating that power-law fits are consistently better 
when rattlers are excluded.

\bibliographystyle{apsrev}